\documentclass[preprint,eqsecnum,tightenlines,nofootinbib,superscriptaddress,prd,showpacs]{revtex4}

 \usepackage{graphicx}
\def\beq{\begin{equation}} \def\eeq{\end{equation}}
\def\bea{\begin{eqnarray}} \def\eea{\end{eqnarray}}

\def\ba{\begin{eqnarray}}
\def\ea{\end{eqnarray}}
\def\be{\begin{equation}}
\def\ee{\end{equation}}

        \def\ee{{\rm e}}

        \def\d{{\rm d}}

       \def\={\mathrel{\widehat\mathalpha{=}}}

\begin{document}

\preprint{\vbox{\baselineskip=12pt \rightline{ICN-UNAM-03/11}
\rightline{hep-th/0311030} }}

\title{Black Holes in de Sitter Space:\\
 Masses, Energies and Entropy Bounds}
\author{Alejandro Corichi}\email{corichi@nuclecu.unam.mx}
\affiliation{Instituto de Ciencias Nucleares\\
 Universidad Nacional Aut\'onoma de M\'exico\\
 A. Postal 70-543, M\'exico D.F. 04510,
M\'exico}
\author{Andres Gomberoff}\email{andres@cecs.cl}
\affiliation{Centro de Estudios Cientificos (CECS)\\
Casilla 1469, Valdivia, Chile}


\begin{abstract}
In this paper we consider spacetimes in vacuum general relativity
---possibly coupled to a scalar field--- with a positive
cosmological constant $\Lambda$. We employ the Isolated Horizons
(IH) formalism where the boundary conditions imposed  are that of
two horizons, one of black hole type and the other, serving as
outer boundary, a cosmological horizon. As particular cases, we
consider the Schwarzschild-de Sitter spacetime, in both $2+1$ and
$3+1$ dimensions. Within the IH formalism, it is useful to define
two different notions of energy for the cosmological horizon,
namely, the ``mass" and the ``energy". Empty de Sitter space
provides an striking example of such distinction: its horizon
energy is zero but the horizon mass takes a finite value given by
$\pi /(2\sqrt{\Lambda})$. For both horizons we study their
thermodynamic properties, compare our results with those of
Euclidean Hamiltonian methods and construct some generalized
Bekenstein entropy bounds. We discuss these new entropy bounds and
compare them with some recently proposed entropy bounds in the
cosmological setting.
\end{abstract}
 \maketitle

\section{Introduction}
\label{sec:1}

Black Holes are one of the most enigmatic constructs in present
day physics. On the one hand they are the most simple predictions
of the general theory of relativity \cite{chandra}, and on the
other hand, they are the paradigmatic objects to test possible
quantum theories of gravity; they have been shown to posses
thermodynamical properties \cite{bekenbh,hawkingbh}, whose
statistical origin should be explained by a quantum theory of
gravity \cite{quantumgravity}.

One of the aspects pertaining to gravitational physics that has
gained some recent attention is the behavior of the theory in the
presence of a positive cosmological constant $\Lambda$. Recent
observations suggest that there is a positive cosmological
constant in Nature, and this brings to the picture, among many
others, some features closely related to black holes: the
existence of cosmological event horizons \cite{haw:eh}. These are
causal horizons that exist even in the absence of matter, namely
in de Sitter space. For each geodesic observer (all of which are
equivalent given the homogeneity of the spacetime), there is a
horizon that hides all the events that are inaccessible. A
question that was answered by Gibbons and Hawking is whether
cosmological horizons are subject to the same thermodynamical
interpretation as black hole horizons. They showed that the same
techniques that were applied to the BH were also useful in the
cosmological case: cosmological horizons posses a temperature and
entropy \cite{hawking}.

However, an aspect that was not uniquely defined at the time was
the issue of associating a mass to the cosmological horizon,
including the case of de Sitter space. The issue of finding an
appropriate definition of mass in asymptotically de Sitter
spacetimes is of course, not new. After the original
Gibbons-Hawking construction, several new proposals have appeared
in the past years, some of which involve assigning a {\it
negative} mass to de Sitter spacetime
\cite{hawking,deser:abbott,mass1,vijay,mann,kastor,padma,teitel,GT}.
The root of this particular feature, in the cases where the first
law is used in the definition, is that the surface gravity
$\kappa_c$ of the cosmological horizon is negative. This fact,
together with the usual relation between temperature, mass and
entropy implied by the first law $\delta M=\kappa/8\pi\,\delta A$,
yields the mentioned result. As we shall see in what follows, this
behavior is a general feature of cosmological horizons.

The purpose of this paper is threefold. Firstly, motivated by the
considerations described previously, it is important to see
whether some unanswered questions can find (interesting) answers:
Is it possible to define an energy and mass for a cosmological
horizon? can one have a consistent thermodynamical interpretation
for such horizons? can one define new entropy bounds given the
notion of energy contained in a bounded region? As we shall see,
in this paper we will provide affirmative answers to all these
questions, making use of the tools available from the Isolated
Horizons formalism \cite{ih:prl}. The second motivation of this
paper,  based entirely on considerations from the Isolated
Horizons perspective, is that it is important to know whether the
positive $\Lambda$ case can be appropriately dealt with, and
whether one can learn something new pertaining static solutions
(possibly hairy). Finally, we would like to compare our results
with those coming from Euclidean methods.

The structure of the paper is as follows. In Sec.~\ref{sec:2} we
provide a brief (and incomplete) summary of the Isolated Horizons
formalism and of canonical Euclidean methods needed for the
remainder of the paper. However, a reader not particularly
interested in the formalism can safely skip it and continue to
Sec.~\ref{sec:3} where we deal with the Schwarzschild-de Sitter
solution in $2+1$ dimensions. This example is used to set the
stage for the more interesting case of the Schwarzschild-de Sitter
spacetime in $3+1$ dimensions, the subject of Sec.~\ref{sec:4}. In
this section, two possible normalizations of the relevant vector
field are considered, giving rise to two possible definitions of
horizon mass for the BH. In Sec.~\ref{sec:5} we consider the
cosmological horizon and define both its energy, and its mass.
These two quantities do not coincide given that they can be
interpreted to represent different objects. In this case the
horizon energy can be naturally interpreted as the total energy
contained in the region bounded by the cosmological horizon, while
the mass is more an attribute of the horizon itself.In
Sec.~\ref{sec:6} we make some thermodynamical considerations and
analyze the different entropy bounds existent, from our
perspective. Finally, in Sec.~\ref{sec:7} we summarize and
conclude. In an appendix we re-analyze the treatment of the system
in $2+1$ gravity from the perspective of Euclidean methods


%

\section{Preliminaries: Isolated Horizons and Euclidean Methods}
\label{sec:2}

 In this section, we give a brief review of the
techniques used in the remaining of the sections for extracting
the different dynamical and thermodynamical quantities of the
systems under study. We first revise the Isolated Horizon
Formalism, and then we go onto Euclidean methods. Those readers
familiar with the formalisms or those interested only in the new
mass formulae for the horizons can safely skip this section.

\subsubsection{Isolated Horizons}

In this part, we give a brief review of the Isolated Horizon
Formalism, specially the notions that are used in the remaining of
the sections.

 In recent years, a new framework tailored to consider
situations in which a black hole is in equilibrium (nothing falls
in), but which allows for the exterior region to be dynamical, has
been developed \cite{ih:prl}. This {\it Isolated Horizon} (IH)
formalism is now in the position of serving as starting point for
several applications, from the extraction of physical quantities
in numerical relativity \cite{Dreyer:2002mx} to quantum entropy
calculations\cite{abck}. The basic idea is to consider space-times
with an interior boundary (to represent the horizon $\Delta$, or
horizons $\Delta_i$), satisfying quasi-local boundary conditions
ensuring that the horizon remains `isolated'. Although the
boundary conditions are motivated by geometric considerations,
they lead to a well defined action principle and Hamiltonian
framework. Furthermore, the boundary conditions imply that certain
`quasi-local charges' $Q_i$, defined at the horizon $\Delta$,
remain constant `in time', and can thus be regarded as the
analogous of the global charges defined at infinity in the
asymptotically flat context. The isolated horizons Hamiltonian
framework allows to define the notion of {\it Horizon Mass}
$M_\Delta$, as function of the `horizon charges'.

In the Einstein-Maxwell and Einstein-Maxwell-Dilaton systems
considered originally \cite{ih:mech,ac:dil}, the horizon mass
satisfies a Smarr-type formula and a generalized first law in
terms of quantities defined exclusively at the horizon (i.e.
without any reference to infinity). The introduction of non-linear
matter fields like the Yang-Mills field brings unexpected
subtleties to the formalism\cite{cns}. However, one still is in
the position of defining a Horizon Mass, and furthermore, this
Horizon Mass satisfies a first law. The formalism accepts a
cosmological constant without further modifications
\cite{ih:mech}.

A (weakly) isolated horizon $\Delta$ is a non-expanding null
surface generated by a (null) vector field $l^a$. The IH boundary
conditions imply that the acceleration $\kappa$ of $l^a$
($l^a\nabla_al^b=\kappa l^b$) is constant on the horizon $\Delta$.
However, the precise value it takes on each point of phase space
(PS) is not determined a-priori. On the other hand, it is known
that for each vector field $t^a_o$ on space-time, the induced
vector field $X_{t_{o}}$ on phase space is Hamiltonian if and only
if there exists a function $E_{t_{o}}$ such that $\delta
E_{t_{o}}=\Omega (\delta,X_{t_{o}})$, {\it for any vector field
$\delta$ on PS}. This condition can be re-written
 as\cite{afk},
 $\delta E_{t_{o}}=\frac{\kappa_{t_{o}}}{8\pi G}\,\delta a_{\Delta}
+ {\rm work\;\; terms}$. Thus, the first law arises as a necessary
and sufficient condition for the consistency of the Hamiltonian
formulation. Thus, the allowed vector fields $t^a$ will be those
for which the first law holds. Note that there are as many `first
laws' as allowed vector fields $l^a\=t^a$ on the horizon. However,
one would like to have a {\it Physical First Law}, where the
Hamiltonian $E_{t_{o}}$ be identified with the `physical mass'
$M_{\Delta}$ of the horizon. This amounts to finding the `right
$\kappa$'. This `normalization problem' can be easily overcome in
the EM system \cite{ih:mech}. In this case, one chooses the
function $\kappa=\kappa(a_\Delta, Q_\Delta)$ as the corresponding
function for the {\it static} solution with charges $(a_\Delta,
Q_\Delta)$. However, for more complicated matter couplings such as
the EYM system, this procedure is not as straightforward, given
that the space of static solutions might be non-connected. In
these cases, a consistent viewpoint is to abandon the notion of a
globally defined  horizon mass on Phase Space, and to define, for
each branch labeled by $n=n_o$, representing a connected
component of the static sector, a canonical normalization
$t^a_{n_o}$ that yields the Horizon Mass $M^{(n_o)}_{\Delta}$ for
the $n_o$ branch \cite{afk}. The horizon mass takes the form,
 \beq
M^{(n_o)}_{\Delta}(r_\Delta)=\frac{1}{2G}
\int_0^{r_\Delta}\beta_{(n_o)}(r) \, \d r\, ,
 \eeq
  along the $n_o$
``branch" with $r_\Delta$ the horizon radius and where
$\beta(r)=2r\kappa(r)$.

Finally, for any $n$ one can relate the horizon mass
$M^{(n)}_{\Delta}$ to the ADM mass of static black holes.  Recall
first that general Hamiltonian considerations imply that, in the
asymptotically flat context, the total Hamiltonian,
consisting of a term at infinity, the ADM mass, and a
term at the horizon, the Horizon Mass, is constant on every
connected component of static solutions (provided the evolution
vector field $t^a$ agrees with the static Killing field everywhere
on this connected component) \cite{ih:mech,afk}. In the
Einstein-Yang-Mills case, since the Hamiltonian is constant on any
$n$-branch, we can evaluate it at the solution with zero horizon
area.  This is just the soliton, for which the horizon area
$a_\Delta$, and the horizon mass $M_{\Delta}$ vanish.  Hence we
have that $H^{(t_0,n)} = M_{\rm sol}^{(n)}$. Thus, we conclude:
 \beq \label{ymmass}
M^{(n)}_{\rm sol} = M^{(n)}_{\rm ADM} - M^{(n)}_{\Delta}
 \eeq
  on
the entire $n$th branch \cite{cns,afk}. Thus, the ADM mass
contains two contributions, one attributed to the black hole
horizon and the other to the outside `hair', captured by the
`solitonic residue'. The formula (\ref{ymmass}), together with
some energetic considerations \cite{acs}, lead to the model of a
colored black hole as a bound state of an ordinary, `bare', black
hole and a `solitonic residue', where the ADM mass of the colored
black hole of radius $r_{\Delta}$ is given by the ADM mass of the
soliton plus the horizon mass of the `bare' black hole plus the
binding energy \cite{acs}.

\begin{figure}
  \includegraphics[angle=360,scale=.55]{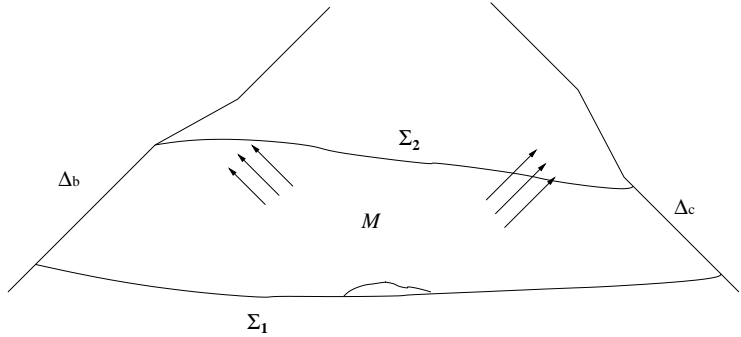}
  \caption{\label{f0}
The action principle for the IH formalism is taken in the region
$M$ of spacetime, bounded by the initial (partial) Cauchy Surface
$\Sigma_1$ and the final surface $\Sigma_2$ and by the two
isolated horizons $\Delta_b$ and $\Delta_c$. As shown in the
figure, some matter and radiation might be present in the region
of spacetime under consideration.}
\end{figure}

In pure Einstein gravity with a positive cosmological constant the
formalism has to be appropriately adapted. The quasi-local
geometrical conditions defining the horizon are not modified. In
the case that we are interested in this paper, namely space-times
with both a black hole and a cosmological horizon, one has to
specify boundary conditions on both horizons (see Fig.~\ref{f0}).
Among the assumptions one has to specify that the region of
interest $M$ is bounded by a {\it black hole type} horizon
$\Delta_b$ and a {\it cosmological type} horizon $\Delta_c$.
 Furthermore, the issue of normalization of the vector
field defining horizon mass is more subtle. In the asymptotically
flat context one always had, in the static sector, a properly
normalized Killing field to be used in the normalization. In the
situation considered here, there is no such privileged asymptotic
vector field and one has to find a new prescription. Now, there is
no asymptotic region, so there is no ADM  mass, but the
Hamiltonian responsible will have (on shell) a term coming from
each horizon. That is, the total Hamiltonian $H_t$ is given by,
\beq H_t=E^{c}_{\Delta}-E^{b}_{\Delta}\label{hamil}
 \eeq
where $E^{c}_{\Delta}$ is the energy associated with the boundary
term at the cosmological horizon, and $E^{b}_{\Delta}$ is the
corresponding one for the black hole horizon.
 Finally, let us remark that a
positive cosmological constant with two horizons had been
considered before \cite{ih:mech}, but the issue of the
normalization was not fully addressed. A complete treatment of
this subject is one of the objectives of this work.

\subsubsection{Euclidean Methods}

One may also use Euclidean methods \cite{gibbons} to deal with
systems having more than one horizon \cite{teitel,GT}. Here one is
interested on the statistical mechanics of the system, which is
considered to be in thermodynamical equilibrium. Because the
various horizons will have, in general,  different temperatures,
the system may be in equilibrium with  one of them only. This has
the same flavor as the isolated horizon formalism in the sense
that one may pick one of the horizons present in the system to
proceed with the analysis.

For Kerr--de Sitter geometries, the idea is to construct an action
principle appropriate to treat either of the two horizons as a
boundary. One then may compute the partition function evaluating
the Euclidean path integral of the system. The partition function
will depend on the quantities fixed on the boundary. The other
horizon is treated as any other point of the manifold, namely, no
field is fixed there. In order for the action to have an extremum
in a regular Euclidean geometry, one fourth of the area of that
horizon must be added to the action principle \cite{Btz}. There is
a freedom in using either horizon as a boundary and the system will be in
thermodynamical equilibrium with the temperature of the second,
regular one.

The problem of  normalization of the timelike killing vector will
be present here as well. Normally, within this viewpoint one would
like to consider the ``observer" to be located at the boundary,
and normalize the time variable such that it corresponds to the
proper time measured by that observer. Technically  this implies
requiring the lapse function to be unity at the boundary. This may
be done, for instance, when dealing with asymptotically flat
spacetimes with boundary at infinity, but fails when considering
(anti--) de Sitter spacetimes.  In the anti-de Sitter case
\cite{HT}, one may normalize the killing vector at infinity
requiring that the associated canonical generators of the AdS
group have the standard normalization (standard structure
constants). In \cite{GT} the de Sitter case was treated in such a
way that, on the one hand,  in the limit of vanishing cosmological
constant the usual flat space normalization was recovered and, on
the other hand, the algebra of charges corresponds to the
continuation of dS to AdS by changing the sign of the cosmological
constant. In this paper we shall adopt a slightly different
viewpoint.

\section{de Sitter in 2+1 dimensions}
\label{sec:3}

In this section we shall consider $2+1$ dimensional gravity with a
positive cosmological constant. It is known that there are no
black hole solutions in this case, but there are cosmological
horizons. In the literature, it has been useful to consider a de
Sitter spacetime with a point particle at the origin \cite{deser}.
This is the situation that we shall study. Note that in this
section we will consider space-times with only one horizon.

Let us recall the general form of the de Sitter space-time with a
conical defect at a pole, \beq \d
s^2=-\left(\alpha^2-\frac{r^2}{l^2}\right)\d
t^2+\left(\alpha^2-\frac{r^2}{l^2}\right)^{-1}\d r^2+r^2\d \phi^2
\eeq where $\Lambda=1/l^2$ is the positive cosmological constant and
\begin{equation}
\alpha =1-4GE \ ,
\label{conical}
\end{equation}
where  $E$ is the energy of the particle sitting
at the origin $r=0$. This
 energy, computed first in
\cite{deser}, may  also be obtained as the conjugate of the
killing vector
$$
\frac{1}{\alpha}\frac{\partial}{\partial t}.
$$
In appendix A a
treatment within an Euclidean canonical formulation is given. The case of
empty de Sitter is recovered when $E=0$, namely, $\alpha=1$. This
spacetime has a horizon located at,
 \beq
r_\Delta=l\alpha
\eeq
The surface gravity $\kappa$ of the Killing
vector field $\ell=(\partial/\partial t)$, that generates the
cosmological horizon is given by,
 \beq
\kappa=-\frac{r_\Delta}{l^2}=-\frac{\alpha}{l}
 \eeq
However, the vector field $\ell$ is not normalized at the origin
since $\ell\cdot\ell |_{r=0}=-\alpha^2$. We would like to choose
as a normalization for the vector field $\ell$ such that it is
normalized with respect to the privileged observer at $r=0$. From
the Euclidean perspective this is quite natural. The Euclidean
geometry can be arranged to be regular at the cosmological horizon
(fixing the appropriate temperature) but it is always singular at
$r=0$ if $\alpha\neq 1$, therefore, we  chose $r=0$ as the
boundary, and we demand the killing vector to be properly
normalized in there.

Thus, we define the new KVF $\tilde{\ell}:=\ell/\alpha$, such that
$\tilde{\ell}\cdot\tilde{\ell}|_{r=0}=-1$. Thus, the new surface
gravity is given by,
 \beq
\tilde{\kappa}=\frac{\kappa}{\alpha}=-\frac{1}{l}\label{temp3d}
\eeq We would like to argue that the temperature that is assigned
to this normalized KVF is physically relevant since this
corresponds to the temperature that  the privileged observer at
the origin measures. The temperature $|\kappa|/2\pi$ is not
measured by any observer. The first observation is that the
physical temperature one would assign to such family of
space-times, namely \beq
T=\frac{|\tilde{\kappa}|}{2\pi}=\frac{1}{2\pi l}\label{newtemp}
 \eeq
is independent of $E$, that is, it is insensitive to the existence
of the massive particle at the origin. This is the first
unexpected observation of this note.

Let us now consider The first law for Isolated Horizons that reads
in this case \cite{ih2}, \beq \delta E_{\Delta}=\frac{\kappa}{8\pi
G}\,\delta a_{\Delta}
 \label{m0}
 \eeq
Here $E_{\Delta}$ is the horizon contribution to the Hamiltonian
that we are identifying as the horizon energy. It is valid for any
choice of $\kappa$. Using our choice of surface gravity
$\tilde{\kappa}$ given by Eq. (\ref{temp3d}), we can now integrate
the first law to get,
 \beq
E_{\Delta}=-\frac{r_{\Delta}}{4Gl}+E_0=-\frac{1}{8\pi G
l}a_{\Delta}+E_0\ , \label{m2}
 \eeq
where $E_0$ is an integration constant.

The horizon radius, for a point particle given by $\alpha$ is then
$r_{\Delta}=l\alpha$. Thus, the cosmological horizon energy of
such spacetime is,
 \beq
E_{\Delta}=-\frac{\alpha}{4G}  + E_0 \, .\label{mass3d}
 \label{m3}
 \eeq
  Note that with this normalization, the mass of the point particle is
 equal to  $E$ in (\ref{conical}) if we set $E_0=1/4G$,
 so that de Sitter is defined to have zero horizon energy.
 Thus we can compute the horizon energy of the cosmological
 horizon for the point particle spacetime.
\beq
E_\Delta=-\frac{\alpha}{4G}+\frac{1}{4G}=\frac{1-\alpha}{4G}=E
\eeq
That is, as expected, the surface contribution to the
Hamiltonian is measuring the total energy contained within the
horizon (i.e. the particle). In 2+1 gravity we don't expect to
have, in vacuum, more contribution to the energy coming from the
geometry. That is, the total energy should correspond to the
contribution from the point particles in the interior region. As
we shall see in the following sections, such interpretation for
the horizon energy of a cosmological horizon continues to be valid
in 3+1 dimensions.  It is interesting to note that the horizon
Energy is independent of the value of the cosmological constant;
it only cares about the ``mass" of the point particle at the
origin.

The above analysis corresponds to setting $N=1/\alpha$ in Eq.
(\ref{var}) of the Appendix A, and taking the boundary at the
cosmological horizon. Nevertheless, from the Euclidean
perspective, the arguments leading to (\ref{m3})  have a different
origin. In this formulation one does not impose Eq. (\ref{m0}).
Instead, one takes the boundary -- as it is normally done in
asymptotically flat space -- to be the outermost sphere in
Euclidean space, that is, the cosmological horizon. Note however,
that in the Euclidean formalism is much more natural to place the
boundary on the particle, because doing so one can remove the
conical singularity produced by it. The cosmological horizon is
then a regular point, and the system may be considered to be in
equilibrium at its temperature. The two choices for the boundary
represent two different solutions of Euclidean gravity (see
\cite{teitel, GT}), and therefore it is natural they have
different energies. These two energies correspond, in the
terminology of the IH formalism, to the distinction between the
``mass" and ``energy" of the horizon.

Let us now consider the Horizon Mass $M_\Delta$. Intuitively, it
should be closely related to the horizon energy. However, one of
the main points of this paper is to argue that such quantities
represent different physical objects. The 3 dimensional system we
are considering is a good example for showing this fact.

Let us define the first law for the Isolated Horizons Mass as
follows,
 \beq \delta
M_{\Delta}=\frac{|\kappa|}{8\pi G}\,\delta a_{\Delta}
 \label{m1}
 \eeq
This form of the first law is consistent to the one adopted in the
context of the Euclidean formalism, where $|\kappa|$ corresponds
to the temperature of the horizon, whose inverse is a positive
quantity, namely, the period in the Euclidean time.

Using our choice of surface gravity $\tilde{\kappa}$ given by Eq.
(\ref{temp3d}), we can now integrate the first law to get,
 \beq
M_{\Delta}=\frac{1}{8\pi G l}a_{\Delta}+M_0\ , \label{m2b}
 \eeq
where $M_0$ is an integration constant.  Thus, the cosmological
horizon mass of such spacetime is,
 \beq
M_{\Delta}=\frac{\alpha}{4G}  + M_0 \, .
 \label{m3b}
 \eeq
This expression is exactly what is obtained in Appendix A using
canonical Euclidean methods when the boundary is placed at the
particle, leaving the cosmological horizon as a regular point in
the Euclidean manifold satisfying. The system may be interpreted
to be in thermal equilibrium at the temperature of the
cosmological horizon. The first law of thermodynamics, -- which
takes the form (\ref{m1}) in terms of the Lorentzian parameters--
then holds.

The choice of the constant $M_0$ is very important in giving
physical meaning to the quantity $M_{\Delta}$ in (\ref{m3b}).
The normalization we shall
choose is that the horizon mass vanishes when the horizon area
goes to zero. We shall then chose $M_0=0$, so we get
 \beq
M_{\Delta}=\frac{\alpha}{4G}\, .\label{mass3db}
 \eeq
Again, the horizon mass only knows about the particle mass at the
origin and not the cosmological constant.
 When such a particle
is not present, that is, in the case of $2+1$ de Sitter space-time
the cosmological horizon mass is $M^{\rm dS}_{\Delta}=1/4G$.
 Let us now consider the behavior of $M_{\Delta}$ as we increase the
``energy" $E$ of the point mass. We start with $E=0$ that
corresponds to $\alpha=1$ and, as $E$ increases, the parameter
$\alpha$, the horizon radius, and therefore the mass $M_{\Delta}$
decreases. The extreme point is when $4GE$ approaches $1$. In the
limit, the spacetime ``opens-up", the horizon shrinks to zero
radius, and the horizon mass also vanishes (for a discussion of
this spacetime see \cite{cg1}). We see then that our choice of
$M_0$ is justified.

Let us now compare this results with those available in the
literature \cite{vijay,myung,strom}. The first difference is that,
as already mentioned, the temperature is always positive and
constant. This is to be contrasted with the value reported in
\cite{myung,strom} and \cite{vijay}: $T_{\rm ssv}=\alpha/2\pi l$,
which seems to be the standard value in the literature. Regarding
the mass, we can compare our value with that reported in
\cite{vijay}, who report a value of $M_{\rm bbm}=\alpha^2/8G$. In
particular, their value for pure de Sitter is $M^{\rm dS}_{\rm
bbm}=1/8G$. Even when qualitatively similar, our results have
quantitative different values for the horizon mass (\ref{mass3d}),
due to our choice of normalization of the Killing field
(\ref{temp3d}). There is also an important qualitative difference
with \cite{vijay,myung}. The temperature and mass found in our
case are associated to an observer, namely the particle at $r=0$,
whereas the  quantities found in \cite{vijay,myung} refer to
observers in an asymptotic region.

\section{Schwarzschild-de Sitter in $3+1$}
\label{sec:4}

In this section we shall return to the situation in which the
spacetime under consideration possesses two isolated horizons, one
being the black hole horizon and the other the cosmological
horizon.

Let us write down the metric for de Schwarzschild - de Sitter
space (SdS), \beq
 \d s^2=-f^2\d t^2
+f^{-2}\d r^2 +r^2\d \Omega^2
 \eeq
where $f^2=\left(1-\frac{2\mu}{r}-\frac{r^2}{l^2}\right)$,
$l=\sqrt{3/\Lambda}$, and $\mu$ is a ``mass parameter''. If there
were no cosmological constant (and therefore the solution were
Schwarzschild), $\mu$ would correspond to the ADM mass. In the
case that $\mu=0$ we recover (a portion of) de Sitter spacetime.
In these coordinates, the spacetime possesses two horizons, given
by the zeros of $f^2$. The smaller root is called the black hole
horizon $r_b$, and the larger one is the cosmological horizon
$r_c$. The family of spacetimes is parameterized by two numbers,
namely $l$ and $\mu$, or alternatively, the two numbers
$r_{b},r_{c}$. It is convenient to use some relations between $l$
and $\mu$ and the two horizon radii $r_{b},r_{c}$:
 \beq
l^2=r_{b}^2+r_{c}^2+r_{b}r_{c}\qquad;\qquad \mu=\frac{
r_{b}r_{c}(r_{b}+r_{c})}{2l^2}\label{rcandrb}
 \eeq
This family of space-times is well behaved, provided that the
parameter $\mu$ is less than $\mu<l/\sqrt{27}$. In this limit, the
two parameters $r_{b}$ and $r_{c}$ approach each other, and $r$
fails to be a good coordinate. In this limit, the spacetime is
known as the ``Nariai solution" \cite{nairai,Podolsky:1999ts}. The
region between the two horizon becomes a ``tube" with $S^2$
sections of constant area, bounded by the two horizons of the same
area $A_{\rm m}=(4\pi/3)\,l^2$.

\begin{figure}
  \includegraphics[angle=270,scale=.50]{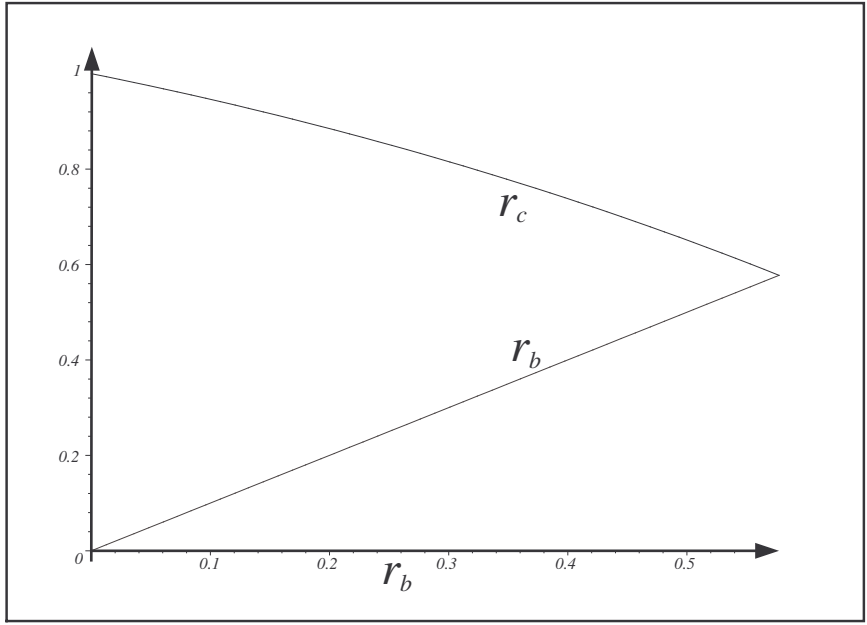}
  \caption{\label{f2}
  The cosmological horizon (geometric) radius $r_c$ and the BH radius
  $r_b$ are plotted as
  functions of  $r_b$. The $y$ axis corresponds to the
  de Sitter limit where there is no black hole ($r_b=0$) and the cosmological
  radius $r_c=l$. The Nariai limit is when both radii coincide.
   In this figure, the parameter $l$ is set to $l=1$.}
\end{figure}

\subsection{The standard normalization}

\label{sn}

In this case, the vector field that we take in order to assign a
surface gravity is $\xi=\partial/\partial t$. This choice is
motivated by the analogy with asymptotically flat solutions and
AdS where this vector has some special properties. For this
choice, the surface gravity is given by \beq
\kappa=\frac{1}{2}(f^2)^\prime|_h=\frac{1}{2}\left(\frac{1}{r_h}-
\frac{3r_h}{l^2}\right) \label{kappaold}
 \eeq
where we denote by $r_h$ the horizon geometrical radius (i.e.
$a_h=4\pi r_h^2$). It turns out that in the black hole horizon
$r_b$, the surface gravity is positive, and in the cosmological
horizon $r_c$, it is negative.

If we now consider this normalization, the IH formalism tells us
that, on each horizon, the first law is valid,
 \beq
  \delta M_\Delta=\frac{|\kappa_h|}{8\pi}\,\delta a_{\Delta}\,
  .\label{1stlaw}
 \eeq
This is a geometrical identity, valid on both horizons
independently.  Therefore, we can integrate and find the black
hole horizon mass
 \beq
M^b_\Delta(r_b)=\int_0^{r_b}\,r_b^\prime\,\kappa(r^\prime_b) \,\d
r^\prime_b=\mu(r_b) \eeq where we have chosen, as boundary
condition that $M^b_\Delta(r_b=0)=0$. It is easy to see that the
parameter $\mu(r_b)$ is given by:
$$\mu(r_b)=\frac{r_b}{2}\left( 1-\frac{r_b^2}{l^2}\right).$$
This was first obtained  in \cite{teitel} in the Euclidean version
of the theory. The mass of the cosmological horizon, is given,  by
\begin{equation}
M^c_\Delta = -\mu + M_0 \ , \label{cosmass}
\end{equation}
where $M_0$ is an integration constant. In \cite{teitel,GT} $M_0$ was
taken to be equal to zero. Here we will take a different point of
view. Note that in the Nariai limit the two horizons become
undistinguishable (and, in fact, the two corresponding Euclidean
instantons become the same). Hence, we will set the $M_0$ so that
in the Nariai limit $M^b_\Delta=M^c_\Delta$, that is,
\begin{equation}
M_0 = \frac{2l}{\sqrt{27}} \ . \label{mo}
\end{equation}
We therefore conclude that the horizon mass of pure de Sitter is
not zero, but $M_\Delta^{\rm dS} =M_0$.

Note that in the interval of interest, $r_b\in [0,l/\sqrt{3}]$,
$\mu(r_b)$ is a monotonic function of $r_b$. It is interesting to
note that in the extreme, Nariai limit, the ``temperature",
defined as $T=\kappa/2\pi$, vanishes. However, one has to note
that this particular normalization, even when natural from the
viewpoint of the canonical coordinates, does not correspond to any
observer in the region of spacetime that we are considering. Thus,
if we adopt the viewpoint that the Killing field should be adapted
to at least one observer inside the spacetime, then one should
look for a new normalization.

\subsection{The Bousso-Hawking normalization}

The issue of normalization of the time evolution vector field
$t^a$ is a very important one within the isolated horizon
perspective (and the Euclidean approach as well). Let us review
very briefly how it is normally done, in the asymptotically flat
context. Given an IH with charges $Q_i$, then one looks for the
(unique) stationary solution with those charges and look for the
surface gravity of the corresponding KVF $K^a$, that has some
special property. Once one chooses the function $\kappa(Q_i)$, one
extends this function to the whole IH phase space. In the case of
asymptotically flat spacetimes, the standard choice is to select
the Killing field that goes to a unit time translation at
infinity. That is, the one that corresponds to the four-velocity
of an observer at infinity. In a sense, this is the most natural
choice. In our case, we do not have an asymptotic region, and
therefore no preferred observer ``far away''. One has to think of
a new normalization criteria. Fortunately, such normalization is
available, as was suggested by Bousso and Hawking, in the context
of Euclidean instantons \cite{bousso}. The idea is to select the
preferred observer, following the integral curves of the Killing
field, for which the acceleration vanishes. There is a unique
value of $r_g$ for which the field $\partial/\partial t$ is
geodesic. We then normalize the Killing field such that
$K^aK_a|_{r_g}=-1$. It is easy to see that one has to choose,
 \beq
K^a=\frac{1}{f(r_{g})}\,\left(\frac{\partial}{\partial t}\right)^a
\eeq Now, let us denote by $\alpha(\mu, l)=1/f(r_g)$.The
normalized surface gravity is then given by,
 \beq
\tilde{\kappa}_h:=\frac{1}{2\sqrt{1-\left(\frac{27\mu^2}{l^2}
\right)^{1/3}}}\left[ \frac{1}{r_h}-\frac{3r_h}{l^2}\right] \eeq
This surface gravity has some nice properties. In particular, in
the Nariai limit, it does not vanish but approaches a constant
value given by $\kappa_{\rm nairai}=\sqrt{3}/l$.

We are now in the position of computing the black hole horizon
mass, by integrating the first law (\ref{1stlaw}), \bea
\tilde{M}_\Delta(r_b)&:=&\int_0^{r_b}r^\prime_b\,\tilde{\kappa}
(r^\prime_b)\,\d
r^\prime_b\nonumber\\
 &=&
-\frac{1}{2}(l\mu)^{1/3}\sqrt{l^{2/3}-3\mu^{2/3}}\nonumber\\
&{}&+\frac{l\sqrt{3}}{6}
\arcsin\left[\sqrt{3}\left(\frac{\mu}{l}\right)^{1/3}\right]\, .
\label{hmcg} \eea
 This is a function of the black hole horizon $r_b$, since the
 specification of the value of $l$ and $r_b$ fixes the parameter
 $\mu(r_b)={r_b}/{2}\left( 1-{r_b^2}/{l^2}\right)$.
This is the `new black hole mass', constructed from the physically
motivated condition of Bousso and Hawking; it will be the mass
that we shall adopt in the remainder of this paper. We shall also
drop the `tildes' over the surface gravity and mass and simply
refer to them as $\kappa(r_b)$ and $M^b_\Delta$ respectively.


\begin{figure}
  \includegraphics[angle=270,scale=.50]{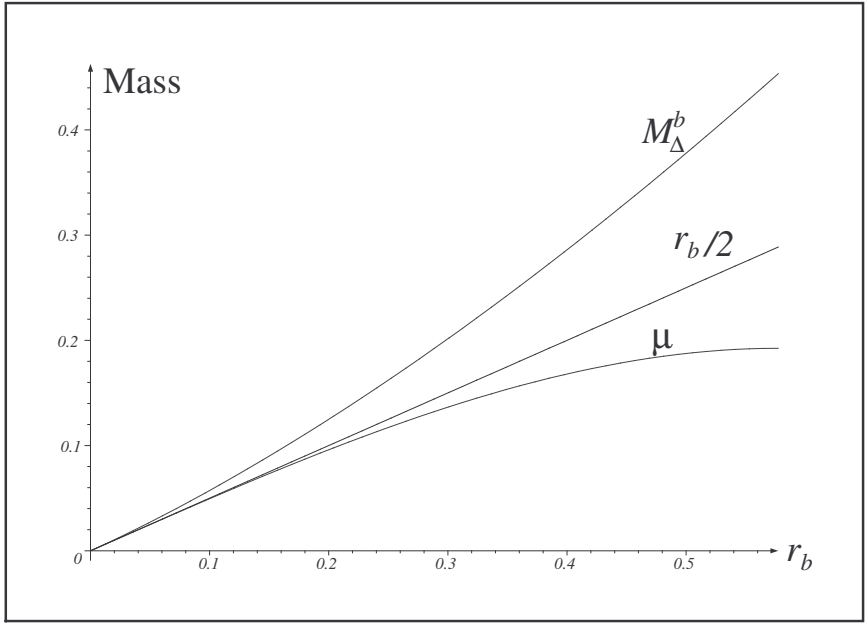}
  \caption{\label{f3}
  The three possible definitions of BH Horizon Mass are drawn as
  function of the BH geometrical radius $r_b$. The line in the middle
  is $r_b/2$, the lower line corresponds to
  $\mu(r_b)$ and the higher line to $M^b_\Delta$.}
\end{figure}

Let us now compare our horizon mass $M^b_\Delta$ with some other
proposals for the black hole horizon mass in the literature. In
the early papers \cite{hawking}, the mass of the black hole was
found to be equal to $\mu$, the ``mass parameter". The
cosmological horizon mass was found to be equal to $-\mu$. This is
also the result found more recently using Euclidean Hamiltonian
techniques \cite{teitel,GT}. It is not difficult to see that this
result is a consequence of using the standard surface gravity
(\ref{kappaold}) associated to the killing time $t$, and of
integrating the first law for both masses, with trivial
integration constants. Some other papers, making use of
quasi-local methods in the manner of Brown-York, find different
expressions for the total mass. See \cite{vijay} and \cite{mann}
for two such approaches, none of which coincides with ours. In a
different approach, using some thermodynamical considerations, the
black hole mass is taken to be equal to $r_b/2$ \cite{padma}. A
comparison of the three possible definitions for the black hole
mass are shown in Fig.~\ref{f3}.

\section{Cosmological Horizon Energy and Mass}
\label{sec:5}

In this section, we shall focus on the cosmological horizon. In
the first part we shall consider the horizon energy coming from
our choice of normalization. Energetic consideration will allow to
say something about more generic dynamical spacetimes. In the
second part, we shall define the cosmological horizon mass. It is
related to the energy but has different properties. Finally, We
discuss the differences between them.

\subsection{Energetics}

In this subsection, we shall consider the Hamiltonian, and as a
result, the energy of the space-time. We shall return to the issue
of the cosmological horizon mass later on.

Recall from Sec.II that the form of the Hamiltonian is given by
Eq.(\ref{hamil}), with two contributions, one from each horizon.
From the general formalism of isolated horizons, each of the
surface terms (accounting for the horizon ``energy") will satisfy
a first law of the form,
 \beq
  \delta E^i_\Delta=\frac{\kappa^i_h}{8\pi}\,\delta a_{\Delta}\,
  .\label{1stlawb}
\eeq
 where $i=b,c$ is a label for the horizons. Now, let us recall
that the sign of the surface gravity $\kappa^i$ depends on the
nature of the horizon; it is positive for the BH horizon and
negative for the cosmological horizon.   With this convention, the
change of the total Hamiltonian of the system is such that
 \beq
\delta H_t=\delta E^{c}_{\Delta}-\delta E^{b}_{\Delta}
 \label{hamil2}
 \eeq

 Let us now discuss the physical motivation for
these choices. Let us assume that we have, as initial condition, a
Schwarzschild de Sitter spacetime, and an inertial observer in the
region between the two horizons with a test mass $m$ (much smaller
than the BH mass), and we throw the test mass across the
cosmological horizon. Then, for a positive change in area of the
cosmological horizon (i.e. when it grows), the change in the
horizon energy $E^c$ is negative $\delta E^c<0$. The change in the
total Hamiltonian $\delta H_t$ will also be negative. Let us now
suppose that the same test mass crosses the BH horizon. In this
case, the BH horizon also grows and therefore $\delta E^{b}>0$.
However, the change in the total Hamiltonian $\delta H_t$ is again
negative. Thus we see that it is natural to regard the Total
Hamiltonian $H_t$ as a measure of the total energy contained
between the horizons. This is very similar to the asymptotically
flat case where the difference in the ADM energy and the BH energy
is given by the total energy radiated through infinity
\cite{ih:mech}. If we follow this reasoning, we should then state
that the quantity,
 \beq
 E^c_{\Delta}-  E^b_{\Delta}=E^{\rm rad}
\eeq
 is equal to the {\it total available energy to be radiated} across both
horizons. Even when we know some important properties for the
cosmological horizon energy $E_{\Delta}^c$, we do not have a
functional form on the full IH phase space. The first law for the
cosmological horizon tells us that $E_{\Delta}^c$ is a function of
$r_\Delta$ only, just as the BH horizon mass (\ref{hmcg}).

In order to find the functional form for the cosmological horizon
energy, it is convenient to consider the case of {\it static}
space-times. We know from the general theory of symplectic
dynamics that the change in the value of the total Hamiltonian on
a static solution is zero. Thus when considering the relation
(\ref{hamil2}) connecting  static solutions we have
 \beq
\delta H_t=\delta E^{c}_{\Delta}(r_c)-\delta E^{b}_{\Delta}(r_b)=0
 \label{hamil3}
 \eeq
Now, it is important to recall that in the static case, namely for
the Schwarzschild-de Sitter family, the cosmological radius $r_c$
and the black hole radius are related by (\ref{rcandrb}). This
means that in Eq. (\ref{hamil3}) one can parameterize the
connected component of the static sector by one parameter, for
instance $r_b$. We then have that for the SdS family, the total
Hamiltonian is a constant,
 \beq
  H_t= E^{c}_{\Delta}(r_c)- E^{b}_{\Delta}(r_b)=C
 \label{hamil4}
 \eeq

\begin{figure}
  \includegraphics[angle=270,scale=.50]{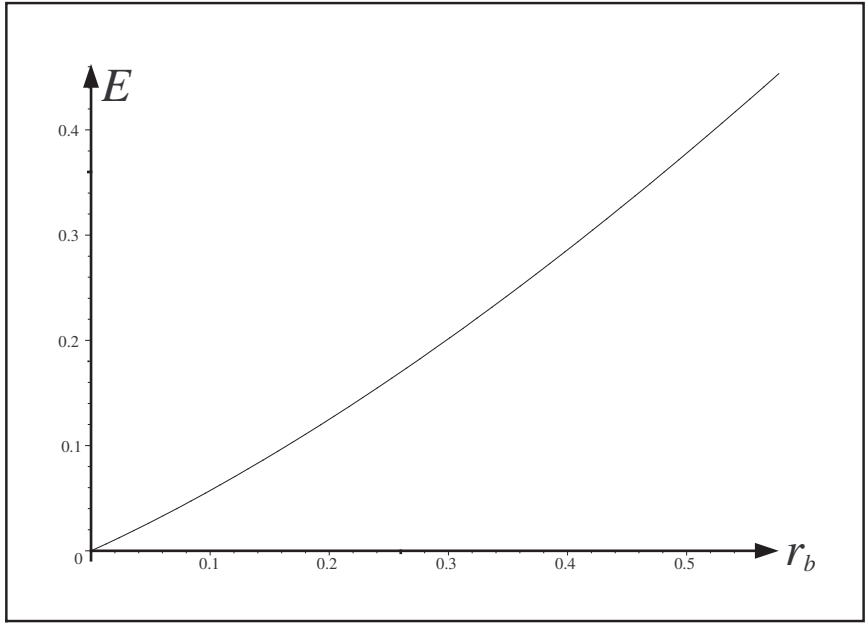}
  \caption{\label{f3b}
The total energy contained in the $SdS$ space-time is plotted as
function of the black hole radius $r_b$. The lower limit $r_b=0$
corresponds to empty de-Sitter space-time where the total energy
is zero.}
\end{figure}

\begin{figure}
  \includegraphics[angle=270,scale=.50]{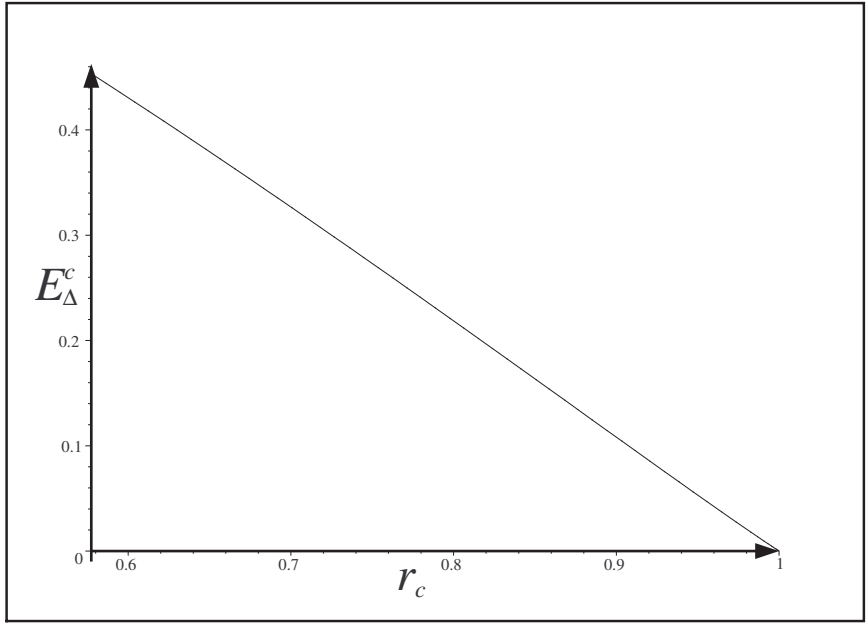}
  \caption{\label{f3bis}
The total energy $E^c_\Delta(r_c)$ contained in a space-time with
a cosmological isolated horizon as an outer boundary is plotted as
function of the cosmological horizon radius $r_c$. The $r_c=1$
limit corresponds to empty de-Sitter where the total energy is
zero.}
\end{figure}

Since $C$ is a constant, we can try to evaluate the LHS of the
equation in the limit when $r_b=0$. In this case, there is no
Black Hole and therefore the value of the total Hamiltonian
corresponds to the total energy contained inside the (vacuum) de
Sitter horizon. We shall assume that this energy is zero
\footnote{This choice is justified, by analogy with the case of
empty anti-de Sitter spacetime where the total ADM energy is zero
\cite{asht-das} , even when there might be some ``vacuum energy"
due to the cosmological constant.}. Therefore, we have that as
functions of $r_b$, both horizon energies have the same value.
This is true in the static family where the value of the BH
horizon radius uniquely determines the value of the cosmological
horizon radius (See Fig.~\ref{f3b}). Using this relation valid for
the static family, we arrive at the cosmological horizon energy
$E_\Delta^c(r_c)$ given by,
 \bea
E^c_\Delta(r_c)&=&-\frac{1}{2}(l\mu)^{1/3}\sqrt{l^{2/3}-3\mu^{2/3}}\nonumber\\
&{}&+\frac{l\sqrt{3}}{6}
\arcsin\left[\sqrt{3}\left(\frac{\mu}{l}\right)^{1/3}\right]\, .
\label{enercosmo}
 \eea
with $\mu(r_c)={r_c}/{2}\left( 1-{r_c^2}/{l^2}\right)$. Thus, as a
function of its area, the cosmological horizon energy is a
decreasing function; It takes its maximum value when the horizon
area is the smallest, namely in the Nariai limit, and it decreases
monotonically until it reaches zero at the de Sitter limit (see
Fig.~\ref{f3bis}). It is important to note that the expression
(\ref{enercosmo}), as a function of the cosmological horizon
radius $r_c$ will be the total horizon energy {\it on the entire
phase space}. That is, on a generic spacetime containing a
cosmological isolated horizon of radius $r_c$ ---independently of
whether a black hole is present or not--- the total energy
contained within will be given by the expression
(\ref{enercosmo}).

Two remark are in order. i) Throughout this paper, we are assuming
that, just as in the static SdS case where $r$ is a good
coordinate in the region of interest, in the full IH phase space
the cosmological horizon is larger than the black hole horizon.
Assuming the validity of our energy formula (\ref{enercosmo}) then
implies that a {\it $\Lambda>0$ Penrose conjecture} is valid
\cite{penrose}. ii) Given a cosmological constant $l$, there is a
maximum bound on the amount of energy that can be contained within
a cosmological horizon given by the Nariai limit of the
cosmological horizon energy, namely $E_{\rm max}=l\pi\sqrt{3}/12$.

\subsection{Cosmological Horizon Mass}

In this part we shall consider the cosmological horizon mass.  The
first thing one needs to specify are ``boundary conditions" for
integrating the first law, which is valid for both types of
horizons. The other input, namely the choice of the preferred
normalization of the surface gravity, is already at our disposal,
so the only remaining input comes from the choice of ``initial
data" when integrating the first law. Let us be more specific. In
the case of black hole horizons (for a given theory, which means
fixing $l$), we have a one parameter family of such spacetimes,
labelled by $\mu$, such that for $\mu=0$ there is no black hole,
and therefore, no horizon. At the same time, the point $\mu=0$
represents the de Sitter universe with a cosmological horizon
$r_{\rm dS}=l$. For small values of $\mu$ (in $l$ units), one
expects that the horizon mass behaves as the horizon mass of a
Schwarzschild BH would, given that in the vicinity of such a
horizon, the geometry ``looks like" a Schwarzschild BH. This
expectation is indeed satisfied since the horizon mass
(\ref{hmcg}) behaves as $\tilde{M}_\Delta(r_b)\approx \mu \approx
r_b/2 $, when $\mu\ll l$. As we increase the value of $\mu$, the
black hole horizon grows, and the cosmological horizon
``shrinks'', up to the Nariai point in which both horizons have
the same size.

 What is then the value of the
Mass that we shall assign to the cosmological horizon? The first
observation is that, from the first law, the variation of the
cosmological energy and mass should be related. Thus, let us
define the cosmological horizon mass $M_{\Delta}^c$ to be such
that $\delta M_{\Delta}^c=-\delta E_\Delta^c=|\kappa|/8\pi\,\delta
a_\Delta$. This proposal is motivated by the following
consideration. Let us recall the situation considered before where
an observer in the region between both horizons, would through a
test mass $\Delta m$ into the cosmological horizon. We know that
the horizon will grow, so $\delta a_\Delta>0$. The change in
cosmological horizon energy will be negative, since this quantity
measures the total amount of energy contained in the partial
Cauchy surface ``inside the horizon". However, one could expect
the cosmological horizon mass to increase by such process, just as
it happens for BH horizons where a test mass falling in decreases
the energy outside the BH but increases the BH mass by the same
amount.

The next question is then about the constant $M_0$ that relates
both functions as
$$M_{\Delta}^c=M_0- E_\Delta^c.$$
 The prescription we
shall put forward in this paper is the following: let us consider
the plot of $M_\Delta^b$ and $M_\Delta^c$ as functions of $r_b$
(or alternatively, $\mu$). We know that the black hole mass
$M_\Delta^b(\mu)$, is a monotonically increasing function of
$\mu$, starting at zero, and reaching its maximum value at the
Nariai point $r_{\rm max}=l/\sqrt{3}$, of $M_\Delta^b(r_{\rm
max})=l\pi\sqrt{3}/12$. We shall now assume that the cosmological
horizon mass is also positive, and reaches its maximum value when
the horizon is largest, that is, in the pure de Sitter case. Then,
the ``continuity proposal'' is that both masses coincide at the
Nariai limit. That is, $M_{\Delta}^c(r_c)=M_{\Delta}^b(r_b)$ when
$r_b=r_c$. As already mentioned in Section \ref{sn}, this is consistent with
the Euclidean formulation, where the solution with the boundary
at the cosmological horizon is not distinguishable from the solution
with the boundary taken at the black hole horizon in the Nariai limit.

\begin{figure}
  \includegraphics[angle=270,scale=.50]{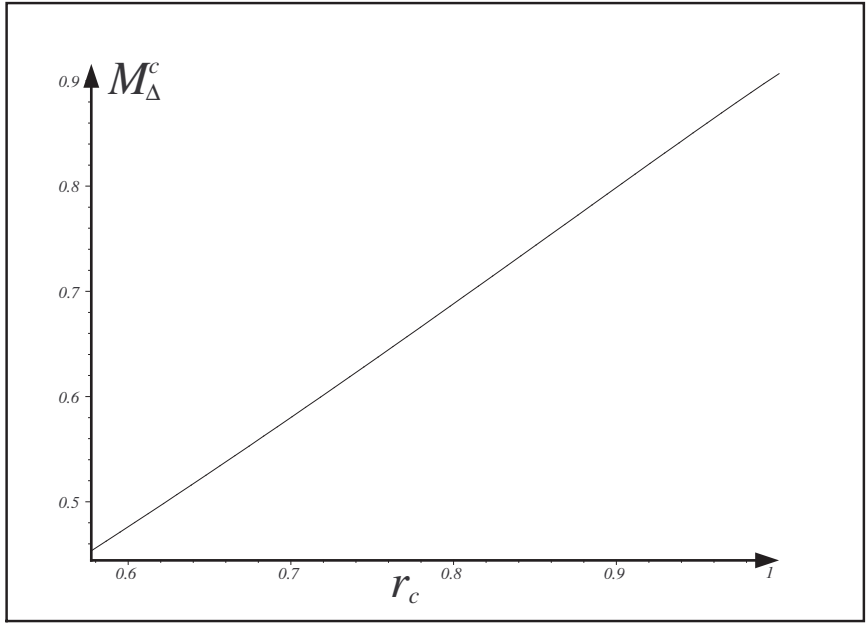}
  \caption{\label{f4}
  The Cosmological Horizon Mass $M^c_\Delta$ is shown as function of the
  cosmological horizon radius $r_c$.}
\end{figure}

Thus, given the symmetries of the surface gravity
$\tilde{\kappa}$, the plot of the masses, as functions of $r_b$ is
symmetric with respect to the horizontal line at
$l\pi\sqrt{3}/12$. It is then straightforward to arrive at the
conclusion that the integration constant $M_0$ is precisely the
horizon mass of pure de Sitter and is given by,
 \beq
M_0 = M_\Delta^{\rm dS} := \frac{\pi\,l}{2\sqrt{3}}\, .
 \eeq

 This is the second observation of this paper. Note that with this
 proposal, in the static case the sum of the masses --as functions of $r_b$--
 is a constant:
 \beq
 M_\Delta^b(\mu)+M_\Delta^c(\mu)={\rm const.}=M_0=\frac{\pi\,l}{2\sqrt{3}}
\label{sumass}
  \eeq

Let us now consider the general dynamical case, where there might
be radiation escaping through the horizons. We know from the
previous part that the total energy to be radiated is given by the
difference in horizon energies, which in terms of horizon masses
is,
 \beq
 E^{\rm rad}=E^c_\Delta-E^b_\Delta= M_\Delta^{\rm dS}-(
M_\Delta^b + M_\Delta^c)
 \eeq

\section{Some Considerations}
\label{sec:6}

This section has three parts. In the first one, we discuss the
thermodynamics of the horizons, in view of our previous
discussion. In the second part we review different entropy bounds
that have been considered in the literature and propose new ones,
motivated by our definitions of energy and mass for black holes
and cosmological horizons. In the last part, we discuss the
possibility of applying the `bound state' model of hairy black
holes when a positive cosmological constant is present.

\subsection{Thermodynamics}

The first law of horizon mechanics (\ref{1stlaw}), applied to each
horizon suggests that the area can be regarded as entropy and
surface gravity as temperature (with the standard coefficients).
This expectation is realized both by the semiclassical
considerations of Gibbons and Hawking \cite{hawking}, by the
Euclidean methods of \cite{gibbons}, and more recently by
statistical computations within different approaches to quantum
gravity \cite{vafa,abck}. For spacetimes with two horizons, recent
results based on Euclidean methods have shown that indeed the BH
and cosmological horizons are subject to thermodynamic
considerations \cite{teitel,GT}. Let us now see that contrary with
the earlier treatments of the subject \cite{hawking}, we have
consistent relations (i.e. signs) relating all the parameters. Let
us focus on the cosmological horizon given that the black hole
horizon can be treated in the standard manner. Let us recall that
we made a distinction between the horizon mass and energy: the
energy $E_\Delta$ was associated to the total energy content in
the (region of the) spacetime, whereas the mass $M_\Delta$ is a
quantity associated the horizon itself. If we want to make
some thermodynamical considerations regarding the horizon, it is
natural to regard the Mass as the quantity entering the first law
of thermodynamics,
 \beq
\delta M_{\Delta}^c=\frac{|\kappa|}{8\pi}\,\delta a_\Delta
 \eeq
The temperature we would associate to such a horizon would then be
\[
T_c=\frac{\hbar |\kappa_c|}{2\pi}
\]
which is a positive quantity. It is important to note that this
relation is valid, independently of the choice of $\kappa(r_h)$.
Had we chosen the usual normalization \cite{hawking,teitel,GT}, we
would have found  different  values for temperature and mass, but
the same qualitative behavior, namely, positive temperature and
mass. Note that the fact that we get positive masses for both
horizons is a consequence of the basic requirements that we have
imposed, namely that the mass of the horizon grows when the area
increases (positive temperature), that the Horizon mass vanishes
in the zero area limit and that in the Nairai limit both masses
coincide.

\subsection{Entropy bounds}

In recent years, there has been an increasing interest in the so
called ``entropy bounds" in the presence of a cosmological
constant \cite{boussorev,boussoD}. In particular, as consequence
of the generalized second law applied to cosmological horizons,
there is the entropy D-bound that states that the entropy of
matter $S_{\rm m}$ allowed inside the cosmological horizon, at a
given initial time $t_i$, is bounded by \cite{boussoD},
 \beq
S_{\rm m}\leq \frac{1}{4}(A_0-A_c)\, ,\label{d-bound}
 \eeq
where $A_0=4\pi l^2$ is the area of the cosmological horizon of
empty de Sitter, and $A_c$ is the area of the cosmological horizon
when the matter is present at time $t_i$ (which is smaller than
$A_0$). In particular, when a BH is the ``matter" considered, the
bound is satisfied, and it is saturated in empty de Sitter space,
that is, the matter entropy has to vanish in that case
\cite{boussoD}. Bousso has also compared the D-bound to the
holographic entropy bound applied to the cosmological horizon
area,
\beq
 S\leq\frac{A_c}{4}\ .\label{holbound}
\eeq

There is also the original Bekenstein bound \cite{beken} that puts
some limits on the entropy of a system of energy $E$ contained in
a region of ``size $R$",
 \beq
  S_{\rm m} \leq 2\pi\,E\,R\label{bekeb}
\eeq
 In the case of gravitational systems with a positive
cosmological constant, Bousso has suggested that, instead of the
energy $E$ of the system (for which he does not have an
expression), one can replace it by its ``gravitational radius"
$R_{\rm g}$, in such a way that the entropy bound looks like,
 \beq
 S_{\rm m} \leq \pi\,R_{\rm g}\,R\label{bekebousso}
\eeq
 Bousso has argued that these two entropy bounds agree for
large dilute systems in de Sitter space \cite{boussorev}, and that
it coincides with the D-bound in the de Sitter limit.

In our case, we {\em do} have a notion of energy for a
gravitational system with an outer boundary given by a
cosmological horizon, namely  $E^c_{\Delta}(r_c)$. In particular,
when the ``matter'' is a BH in de Sitter, one can ask whether an
entropy bound like (\ref{bekeb}) is valid, without the necessity
of introducing a radius. The expression we would like to propose
when there is a BH present is given by,
\beq
 S_{\rm bh}\leq 2\pi E^b_{\Delta}(r_b)\,r_c\label{entconj}
\eeq
 where the matter system to be considered inside de Sitter is
a black hole of area $A_b=4\pi r_b^2$, with energy equal to
$E^b_{\Delta}(r_b)$, and the maximum size of the system is assumed
to be $r_c$, the cosmological radius (which is smaller than
$r_0=l$, the de Sitter radius). Then, in order to test the
validity of (\ref{entconj}), let us rewrite it as,
$$
r_b\leq\frac{M^b_{\Delta}(r_b)}{(r_b/2)}\,r_c
$$
Now, it is clear from Fig.~\ref{f2} that the term on the RHS
multiplying $r_c$ is always larger than one, and $r_c \geq r_b$.
Therefore, the inequality is always satisfied and is saturated in
the $r_b\to 0$, de Sitter limit. In the general case of a unique
cosmological horizon of radius $r_c$, the modified Bekenstein
bound reads \beq
 S_{\rm m}\leq 2\pi E^c_{\Delta}(r_c)\,r_c\label{entconj2}
\eeq

Let us now consider a more stringent condition, where we assume
that we have a black hole and some matter in the interior region.
We have to consider the entropy of both the BH and the matter.
The condition now reads,
 \beq
 S_{\rm bh}+S_{\rm m}\leq 2\pi E^c_{\Delta}(r_c)\,r_c\label{entconj3}
 \eeq
which can be written as,
 \beq
S_{\rm m}\leq \pi (2E^c_{\Delta}(r_c)\,r_c- r^2_b)\label{entnew2}
 \eeq

Now recall that $E^c_{\Delta}(r_c)$ is equal to the Horizon mass
$M^b_\Delta(r_b)$ of the corresponding BH in SdS and that
$2M_\Delta^b(r_b)>r_b$. Thus, the new entropy bound
(\ref{entnew2}) is well defined. For the case of pure de Sitter,
$r_b=0$, $E^c_{\Delta}(r_c)=0$ and therefore the bound is
saturated by $S_{\rm m}=0$ (just as the D-bound).

\begin{figure}
  \includegraphics[angle=270,scale=.50]{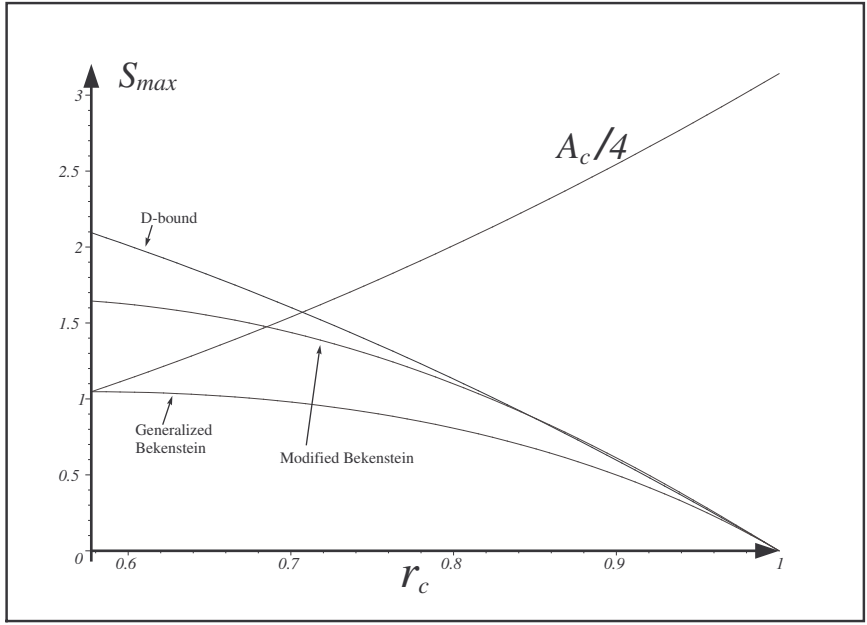}
  \caption{\label{f5}
  Several entropy bounds are shown as functions of the cosmological radius
  $r_c$. For a discussion see the main text.}
\end{figure}
Let us now compare the different entropy bounds that we have
introduced. In Fig.~\ref{f5} we have plotted four of these bounds
as functions of the cosmological horizon radius $r_c$. Three of
them are given by a decreasing function and only one, the
holographic bound (\ref{holbound}), is given by an increasing
function of $r_c$. Of the remaining bounds, the less restrictive
is the D-bound, and the most restrictive is the `generalized
Bekenstein bound' as written by Bousso (\ref{bekebousso}). The
modified Bekenstein bound proposed in this paper (\ref{entconj2})
corresponds to the intermediate line in Fig.~\ref{f5}. In
\cite{boussoD} Bousso has argued that (\ref{bekebousso})
represents an interpolation between the Holographic bound and the
D-bound and can be regarded a valid generalization of the (flat)
Bekenstein bound to cosmological settings. We would like to argue
against this interpretation. First, one should notice that the
Bousso bound (\ref{bekebousso}) was derived in the small $r_b$
limit, and therefore there is no contradiction with it not being
valid near the large BH radius, ``Nairai limit". In fact we can
understand the departure between both bounds for `large' $r_b$
from the fact that in this region, the Horizon Energy
$E^c_{\Delta}(r_b)$ and (one half of) the ``gravitational radius"
$r_b$ depart from each other. That is, the `gravitational radius'
fails to be a good measure of the horizon mass for large black
holes.

\begin{figure}
  \includegraphics[angle=270,scale=.50]{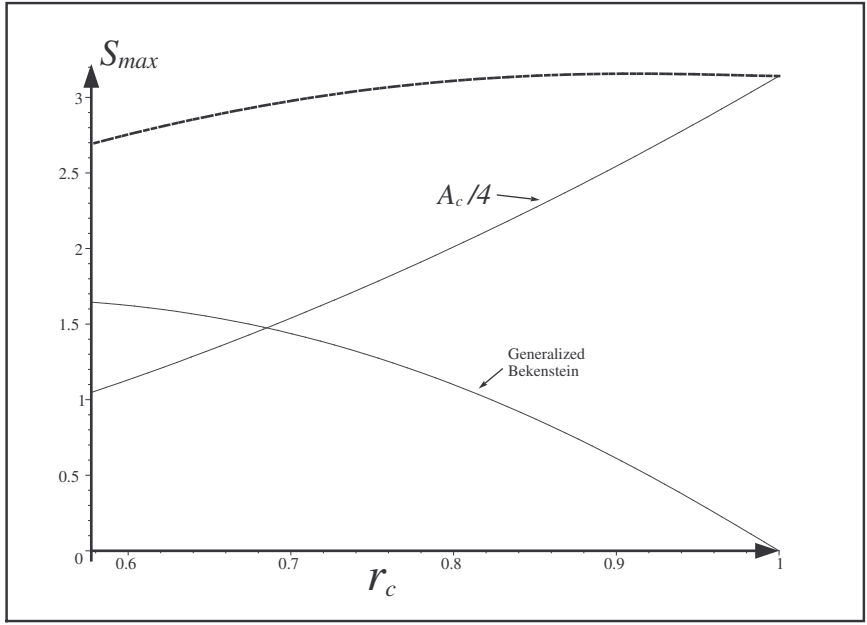}
  \caption{\label{f6}
  The generalized Bekenstein and the Holographic bound, together
  with the sum of them (dotted curve) are shown as functions of
  the cosmological radius
  $r_c$.}
\end{figure}

Second, we would like to argue against the validity of the
holographic entropy bound when applied to the cosmological
horizon, as presently interpreted. For that, let us recall the
argument for assigning entropy to a black hole. If we have some
matter outside the BH, with some entropy $S_m$, and throw the
matter across the horizon, the entropy in the exterior region,
where the `observer' is, will decrease. The area of the BH
increases and the total balance of entropy is saved by associating
entropy to the area of the BH horizon. The standard interpretation
is that the quantity $A/4$ {\it is} the entropy that the external
observer assigns to the horizon and in a sense could be though as
providing some information about the causally disconnected region
inside the horizon. In the case where there is a cosmological
horizon, the observer is in the {\it interior region}, that is, in
the region $r<r_c$. If one throws some matter with a certain
amount of entropy across the cosmological horizon, the entropy in
the observer's region will decrease, the horizon will grow and
therefore the quantity $A/4$ will increase. Thus, in analogy with
the BH case, one is forced to assign the entropy $S_{\rm hor}=A/4$
to the cosmological horizon as seen {\it from the inside}. This
also means that this quantity is giving a measure of the
information contained ``everywhere else" outside the observer's
region. Thus, the holographic bound can not be a bound for the
entropy {\it within} the observer's region $r<r_c$. This can also
explain why the Holographic bound (\ref{holbound}) has a different
behavior as the rest of the bounds; it is bounding the entropy of
a different, disconnected region. What one could do with the
holographic bound, in spirit of the holographic principle
\cite{holog}, is to combine it with the generalized Bekenstein
bound to have a ``total" bound on the entropy everywhere: \beq
 S_{\rm tot}\leq S_{\rm  m}+
S_{\rm out}\leq 2\pi E^c_{\Delta}(r_c)\,r_c + A_{c}/4
\label{totbound}
 \eeq
That is, there is a contribution to the total entropy, as seen by
an observer in the region $r_b<r<r_c$, coming from the matter in
the region (including the BH entropy $A_b/4$) and a contribution
from the `outside' matter, bounded by $A_c/4$. From Fig.~\ref{f6}
we see that the total bound (\ref{totbound}) grows with the size
of the cosmological horizon, but reaches a maximum before the de
Sitter limit.

Let us end this section with a remark. We have generalized the
Bekenstein bound (\ref{bekeb}) using our measure of energy
contained inside the cosmological horizon, but we have not
intended (nor claim) to prove that such a bound is valid in the
{\it strong} gravity regime. As far as we know the validity of
such bound has only been proven in the weak gravity limit
\cite{marolf}. On the other hand, the bound has been shown to be
violated when NUT charges are coupled to gravity and the physical
quantities are computed in the asymptotic region \cite{Clarkson}.

\subsection{Black Holes as Bound States}

One of the most successful applications of the isolated horizons
formalism is the physical model of a static, hairy black hole as a
bound state of a soliton and a ``bare" black hole. This model was
motivated by considerations of spherically symmetric static
solutions to the EYM equations \cite{acs}, but it has proved to be
valid also for axisymmetric solutions of EYM \cite{kunz}, EYMH
\cite{higgs} and Born-Infeld \cite{nora} black holes. A natural
question is whether static solutions with a cosmological constant
can be given such an interpretation. Let us recall that in the
asymptotically flat case, the total (ADM) energy of the spacetime
is written as a sum of the energy of a bare black hole (the
Schwarzschild BH in the non-rotating case), the energy of the
soliton and a ``binding energy".  The first observation is that
the analog of the ADM energy corresponds to the energy of the
spacetime inside de cosmological horizon, and the bare black hole
will be the Schwarzschild de Sitter solution. The total energy can
be written as,
 \beq
E^c_{\Delta}(r_b)=M_0(r_b)+M^{\rm sol} + E_{\rm
bin}(r_b)\label{bind}
 \eeq
where the binding energy $E_{\rm bin}$ is given by
$$E_{\rm bin}(r_b)=-[M_0(r_b)-M_{\Delta}(r_b)]$$
As expected, the binding energy should always be negative, and its
absolute value is a decreasing function of $r_b$. That is, larger
black holes are ``less bounded". This is a general feature
observed in other gravity-matter systems \cite{acs}.

Now, let us recall that in the asymptotically flat case for say,
Einstein Maxwell fields, there are no solitons other than
Minkowski spacetime  (by solitons it is normally understood
regular stationary solutions, even when unstable). In that case
the horizon and ADM masses coincide, the soliton mass is zero and
therefore there is no bound state and the binding energy in
(\ref{bind}) vanishes. Our simple vacuum $\Lambda > 0$ case has
the same behavior. When one couples matter and finds non-trivial,
hairy, solutions one is then led to the `bound state' model
\cite{acs}. The form of the first law we have considered (without
a `work term') will be valid for theories with no `gauge
couplings' like a minimally coupled scalar field (with an
arbitrary potential) or a YM field in the pure magnetic sector.
Non-minimal coupling can also be incorporated to the formalism,
both classically and quantum mechanically \cite{ac:nonmin}, where
the main modification is that the entropy also has a contribution
from the scalar field at the horizon. It would be interesting to
explore gravity-matter systems to find such hairy solutions, and
check whether the `bound state' model is satisfied also for
theories in the presence of a cosmological constant. These
investigations will be reported elsewhere.

\section{Discussion and Outlook}
\label{sec:7}

Let us summarize our results. For static solutions with a
cosmological constant in $2+1$ and $3+1$ dimensions, we have found
mass formulae for both black hole and cosmological horizons.
Furthermore, the powerful isolated horizons formalism based on a
Hamiltonian formulation allowed us to extend these results and
establish a formula for the total energy contained inside a
cosmological horizon in the full IH phase space. This formula in
turn led us to propose a generalized Bekenstein bound for the
entropy of the matter content. This entropy bound was compared
with some other bounds available in the literature. The existence
of a first law of horizon dynamics made it possible to infer some
thermodynamical properties for both horizons. These properties are
qualitatively similar to previous treatments, but differ in the
quantitative value for the thermodynamical parameters. In
particular, the choice of proper normalization for the Killing
field on static solutions given by the Bousso-Hawking prescription
provides a physically acceptable definition of the temperature to
be used in the IH formalism and, upon integration, defines the
(isolated) horizon energy and mass, for appropriate choices of
integration constants.


In our treatment, there was a key input that allowed us to
construct different physical quantities. The general theme of this
construction is that the role of particular observers is crucial
to identify certain quantities. More precisely, by requiring that
a preferred observer be the one to assign a temperature to both
horizons lead us to the Bousso-Hawking normalization. Similarly,
by requiring that observers in the region between the horizons be
the ones that assign energy, mass and entropy to both horizons,
allowed us to propose generalized entropy bounds, even for the
strong gravity regime. Again, in the spirit of the holographic
principle, where different observers might have a different
perception of the world, these two quantities, the mass and energy
for the cosmological horizon are different. In one case, the
energy is associated to the total energy contained within the
horizon (i.e. in the region where the observer is), whereas the
horizon mass is a quantity associated to the horizon itself, that
contains some information about the radiation and matter that has
fallen into it. Furthermore, we were able to pose some conjectures
regarding a maximum value of energy contained in a cosmological
horizon and a `Penrose conjecture' with a cosmological constant.


Let us end with some remarks:

\begin{enumerate}

\item 
The results obtained in most of the sections of this paper were
based on the isolated horizons formalism. However, we have made,
when possible, some comparisons with the results obtained with
Euclidean canonical methods. Let us now compare the similarities
and differences of both approaches. Perhaps the most important
feature that these approaches share is the fact that both are
formulated considering the spacetime region between the two
horizons. That is, they do not consider nor have anything to say
about the interior region of the black hole, nor the asymptotic
region. Second, the physical quantities that both formalisms
produce are tied to a particular observer: in the Euclidean
approach one has the liberty of choosing which horizon is left
`free' and which is taken as a boundary. Both formalisms are
subject to the issue of normalization of the Killing field (or
alternatively `time function'). Finally, in both cases one is free
to add a constant when integrating the first law to define an
energy or mass. As we have repeatedly seen, the choice of such
constants involves a good part of the `physics' of the problem.
The main difference between both approaches are obvious but we
shall enumerate them: The IH formalism is intrinsically Lorentzian
since the (null) boundaries are at the forefront of the formalism.
Furthermore, it allows for generic dynamical situations; static
spacetimes are one of an --infinite dimensional-- realm of
possibilities. On the other hand the Euclidean approach is limited
to static spacetimes (admitting Wick rotations), and the boundary
conditions of the null horizon are difficult to track down.

\item 
The Isolated Horizons formalism as originally formulated for
asymptotically flat spacetimes has been very useful in many
applications. One of its main virtues was the possibility of
implementing boundary conditions and extracting physical
predictions at the horizon, where no preferred background
structure is present --such as Poincar\'e invariance at
infinity--, and thus for the strong gravity regime. There was
however some shadow of the (flat) asymptotic conditions through
the normalization of the evolution vector field at infinity, whose
choice affects the values of both the ADM energy at infinity and
the horizon mass. In our situation with two horizons and no
asymptotic region, one is forced to implement a consistent choice
of normalization that does not make use of an asymptotic region.
In this paper the proposal that we have put forward was based on a
preferred geodesic observer, namely the particle sitting at the
origin in the $3D$ case and the static observer (not accelerating
to any of the horizons) in the $4D$ case. Whether these choices
are the physically most adequate still needs to be explored.

\item 
We have based our treatment on the isolated horizons formalism.
this is well defined in a situation like the one depicted in
Fig.~\ref{f0}, where initial data for SdS is specified, together
with a small perturbation $\Delta m$ in the intermediate region (a
scalar field for instance). The spacetime will evolve dynamically,
the matter field will be radiated and eventually will cross the
horizons, thus violating the `isolated' character of the horizons.
Thus, the region of interest in spacetime is the one before the
radiation reaches either of the horizons. In order to treat the
full dynamical situation one would have to apply the `dynamical
horizons' formalism \cite{dynamical}. However, one should note
that the formalism of \cite{dynamical} would have to be modified
slightly to accommodate for the normalization chosen in this
paper.

\item 
In this paper we have avoided mentioning the dS/CFT correspondence
\cite{ds/cft}, and will continue to do so. This is because we have
{\it not} considered, from the very beginning, the asymptotic
region of spacetime. On the contrary, the Isolated horizon
formulation is restricted to the region of spacetime contained
in-between the horizons (see Fig.~\ref{f0}), a region very far
from the asymptotic future where the correspondence is conjectured
to be valid. In this respect, our mass formulas are conceptually
very different to those of \cite{vijay,mann} that were computed
using the Brown-York approach in the asymptotic future.

\item 
We have {\it not} attempted to make a connection between our mass
formulae and entropy with the approach of the so called
Cardy-Verlinde formula \cite{verlinde}, where the mass definition
that is taken is the one given in \cite{vijay} and the mass
formulae of \cite{deser:abbott}. It would be of interest to see
whether that formalism can be made consistent with our choice of
horizon mass.

\end{enumerate}

The mass and energy formulas that we have proposed in this paper
differ from the standard expressions found elsewhere. In order to
settle the validity of our formulas, it is necessary to have
quantitative dynamical evolutions (with a spherically symmetric
scalar field for instance), for which control on the total amount
of initial and radiated energy is possible. We shall report those
findings in a future communication.

\begin{acknowledgments}
We would like to thank  A. Ashtekar, C. Beetle, S. Fairhurst, G.
Mena Marugan, D. Sudarsky and C. Teitelboim for discussions and
comments. This work was supported in part by the CONACyT
(M\'exico) grant J32754-E and DGAPA-UNAM grant IN112401, and by
CONICYT (Chile) grants No.1010449 and 7010449. A.G. acknowledges
institutional support to CECS provided by the Millennium Science
Initiative, Chile, and by Fundaci\'on Andes, Chile. CECS also
benefits from the generous support by Empresas CMPC. A.G. also
acknowledges support from FONDECYT grant 1010446.

\end{acknowledgments}

\appendix

\section{Euclidean $2+1$}
In this appendix we shall consider the $2+1$ system considered in
Sec.~\ref{sec:3} from the perspective of a canonical Euclidean
approach. Further details on the methods used here may be found in \cite{GT}, where
an analogous treatment for (3+1)-dimensional Schwarzschild--de Sitter is given
in its appendix.

 We assume that  the metric  becomes axially symmetric as it
 approaches the boundary,
 with the form
\begin{equation}
\d s^2= N^2 (\gamma_{,\rho})^2 \d t^2 + \d\rho^2 + \gamma^2
\d\phi^2 \ \ \ , \label{boundaryform}
\end{equation}
The re-scaled lapse $N$ and the coefficient $\gamma$ are functions
of $\rho$ only.  The coordinate $\rho$ is defined so that
$\gamma_{,\rho}$ is negative as one approaches the boundary. The
boundary is a surface of  constant $\rho$. We will be interested
in two possible of such choices: a cosmological horizon or a
conical singularity generated by a point particle. It is
convenient to redefine the hamiltonian generator,
\begin{equation}
\tilde{{\cal H}}_{\perp} = -{\cal H}_{\perp}   \gamma_{,\rho}  \ ,
\label{ham}
\end{equation}
so that the associated Lagrange multiplier is $N$,  and the
corresponding term in the hamiltonian is
\begin{equation}
\int \d^3\! x\,(N^{\perp}{\cal H}_{\perp}) =  \int \d^3\!\,
x(N{\tilde{\cal H}}_{\perp}) \ , \label{action}
\end{equation}
where
\begin{equation}
\tilde{{\cal H}}_\perp = \frac{1}{8\pi G}  \gamma_{,\rho}
\left(\gamma_{,\rho\rho} +\frac{\gamma}{l^2}\right)  \ .
\label{cons}
\end{equation}

 The  boundary term in the variation of the Hamiltonian (\ref{action})
 reads,
\begin{equation}
\left. \frac{1}{16\pi G} \int \d\phi \, N \ \delta \!  \left(
\gamma_{,\rho}^{\ 2} +\frac{\gamma^2}{l^2} \right)
\right|^{\rho_2}_{\rho_1} \  . \label{var}
\end{equation}
Here $\rho_1$, $\rho_2$ are the bounds of the coordinate $\rho$.
Note, however, that in the case at hand only one of them will be
considered to be the boundary, where we will set $N=1$, so that
the surface integral at the boundary may be written as
\begin{equation}
-\delta U \ , \label{}
\end{equation}
where the energy $U$ is given by
\begin{equation}
U= \pm \frac{1}{8G}\left(   \gamma_{,\rho}^{\ 2}
+\frac{\gamma^2}{l^2}  \right) \ . \label{up}
\end{equation}
Here the minus sign is taken when the  boundary is at  $\rho_1$
and the plus sign if the boundary is at $\rho_2$. To evaluate
expression (\ref{up}) for the de Sitter geometry with a point
particle, (\ref{}) we first make  the change of coordinates
$r=\alpha l \cos(\rho/l)$ to rewrite it in the form
(\ref{boundaryform}),
\begin{equation}
\d s^2=\alpha^2 \sin^2\left(\frac{\rho}{l}\right)\d t^2 + \d\rho^2
+ \alpha^2 l^2 \cos^2\left(\frac{\rho}{l}\right) \d\phi^2 \ .
\label{dspart}
\end{equation}
[Note that our prescription  requires the metric  to be cast in
the form  (\ref{boundaryform})  at the boundary only, and not
everywhere as it is possible in this case.]  Here $\rho \in
[0,l\pi/2]$, so that $\rho_1=0$ is the cosmological horizon
$r=\alpha l$ and $\rho_2=l\pi /2$ is the location of the particle,
$r=0$. If we choose the cosmological horizon as the boundary, then
a delta function must be added at the location of the point
particle, $r=0$,  in order for the Einstein equations to be
satisfied \cite{deser}, and the energy reads
\begin{equation}
U_h = -\frac{\alpha^2}{8G} + K_h \ .
\end{equation}
Now,  if the boundary is taken at the particle, then the action
principle must be supplemented with one quarter of the area of the
cosmological horizon\cite{Btz}, and
\begin{equation}
U_p = \frac{\alpha^2}{8G} + K_p \ .
\end{equation}
The constants $K_h$, $K_p$ represent the arbitrariness in fixing
the zero point of the energy. Usually one demands that in the
absence of a conical singularity, that is, when $\alpha^2=1$,  the
energy vanishes, so that $K_h=-K_p=1/8G $. An alternative choice
is to ask that the energy that the particle assigns to a zero area
horizon be zero. In that case, one should choose $K_h=0$.

The energy $U$ considered in this appendix is conjugated to the
Killing time $t$. We may also use a different normalization, and set $N=1/\alpha$
on the boundary. In that case we will get the expressions (\ref{m1}), (\ref{m2}),
(\ref{m3})
in the main text.


\begin{thebibliography}{99}

\bibitem{chandra} S.~Chandrasekhar,
{\it The Mathematical Theory Of Black Holes}, (Oxford U. Press,
1985).


\bibitem{bekenbh}  J.\ D.\ Bekenstein, Black holes and entropy,
{\it Phys.\ Rev.} {\bf D7}, 2333-2346 (1973); J.\ D.\ Bekenstein,
Generalised second law of thermodynamics in black hole physics,
{\it Phys.\ Rev.} {\bf D9} 3292-3300 (1974).

\bibitem{hawkingbh} S.W. Hawking, ``Particle creation by black
holes", Commun. Math. Phys. {\bf 43}, 199 (1975).

\bibitem{quantumgravity} See for instance: G.~T.~Horowitz,
``Quantum gravity at the turn of the millennium,''
arXiv:gr-qc/0011089;
S.~Carlip, ``Quantum gravity: A progress report,'' Rept.\ Prog.\
Phys.\  {\bf 64}, 885 (2001) [arXiv:gr-qc/0108040].


\bibitem{haw:eh} S.W. Hawking, ``The event horizon", in {\it Black
Holes}, eds DeWitt and DeWitt (Gordon and Brach, 1973).

\bibitem{hawking} G.~W.~Gibbons and S.~W.~Hawking,
``Cosmological Event Horizons, Thermodynamics, And Particle
Creation,'' Phys.\ Rev.\ D {\bf 15}, 2738 (1977).

\bibitem{deser:abbott}
L.~F.~Abbott and S.~Deser, ``Stability Of Gravity With A
Cosmological Constant,'' Nucl.\ Phys.\ B {\bf 195}, 76 (1982).


\bibitem{mass1}
K.~Nakao, T.~Shiromizu and K.~Maeda, ``Gravitational mass in
asymptotically de Sitter space-times,'' Class.\ Quant.\ Grav.\
{\bf 11}, 2059 (1994);
T.~Shiromizu, D.~Ida and T.~Torii, ``Gravitational energy, dS/CFT
correspondence and cosmic no-hair,'' JHEP {\bf 0111}, 010 (2001)
[arXiv:hep-th/0109057].
R.~G.~Cai, Y.~S.~Myung and Y.~Z.~Zhang, ``Check of the mass bound
conjecture in de Sitter space,'' Phys.\ Rev.\ D {\bf 65}, 084019
(2002) [arXiv:hep-th/0110234].



\bibitem{vijay} V.~Balasubramanian, J.~de Boer and D.~Minic,
``Mass, entropy and holography in asymptotically de Sitter
spaces,'' Phys.\ Rev.\ D {\bf 65}, 123508 (2002)
[arXiv:hep-th/0110108].

\bibitem{mann} A.~M.~Ghezelbash and R.~B.~Mann,
``Action, mass and entropy of Schwarzschild-de Sitter black holes
and the de Sitter/CFT correspondence,'' JHEP {\bf 0201}, 005
(2002) [arXiv:hep-th/0111217].

\bibitem{kastor} D.~Kastor and J.~Traschen,
``A positive energy theorem for asymptotically deSitter
spacetimes,'' Class.\ Quant.\ Grav.\  {\bf 19}, 5901 (2002)
[arXiv:hep-th/0206105].

\bibitem{padma} T.~Padmanabhan,
``Classical and quantum thermodynamics of horizons in spherically
symmetric spacetimes,'' Class.\ Quant.\ Grav.\  {\bf 19}, 5387
(2002) [arXiv:gr-qc/0204019];
T.~Padmanabhan, ``Thermodynamics and / of horizons: A comparison
of Schwarzschild,  Rindler and de Sitter spacetimes,'' Mod.\
Phys.\ Lett.\ A {\bf 17}, 923 (2002) [arXiv:gr-qc/0202078].

\bibitem{teitel}C.~Teitelboim, ``Gravitational Thermodynamics of Schwarzschild-de
Sitter Space,'' arXiv:hep-th/0203258.

\bibitem{GT}
A.~Gomberoff and C.~Teitelboim, ``de Sitter black holes with
either of the two horizons as a boundary.'' Phys. Rev. {\bf D67},
104024 (2003). arXiv:hep-th/0302204;



\bibitem{ih:prl}
A.~Ashtekar, C.~Beetle, O.~Dreyer, S.~Fairhurst, B.~Krishnan,
J.~Lewandowski and J.~Wisniewski, ``Isolated horizons and their
applications,'' Phys.\ Rev.\ Lett.\  {\bf 85}, 3564 (2000)
[arXiv:gr-qc/0006006].

\bibitem{Dreyer:2002mx}
O.~Dreyer, B.~Krishnan, D.~Shoemaker and E.~Schnetter,
``Introduction to Isolated Horizons in Numerical Relativity,''
Phys.\ Rev.\ D {\bf 67}, 024018 (2003) [arXiv:gr-qc/0206008].

\bibitem{abck} A.~Ashtekar, J.~Baez, A.~Corichi, K.~Krasnov.
Quantum geometry and black hole entropy. \textit{Phys.\ Rev.\
Lett.} \textbf{80} 904-907 (1998) [arXiv:gr-qc/9710007];
 A. Ashtekar, A. Corichi and K.
Krasnov,  Isolated horizons: the classical phase space,
\textit{Adv.\ Theor.\ Math.\ Phys.} \textbf{3} 418-471 (1999)
[arXiv:gr-qc/9905089];
 A.~Ashtekar, J.~Baez, K.~Krasnov. Quantum
geometry of isolated horizons and black hole entropy,
\textit{Adv.\ Theo.\ Math.\ Phys} \textbf{4}, 1-95 (2000)
[arXiv:gr-qc/0005126].


\bibitem{ih:mech}
A.~Ashtekar, C.~Beetle and S.~Fairhurst, ``Mechanics of Isolated
Horizons,'' Class.\ Quant.\ Grav.\  {\bf 17}, 253 (2000)
[arXiv:gr-qc/9907068].

\bibitem{ac:dil}
A.~Ashtekar and A.~Corichi, ``Laws governing isolated horizons:
Inclusion of dilaton couplings,'' Class.\ Quant.\ Grav.\  {\bf
17}, 1317 (2000) [arXiv:gr-qc/9910068].

\bibitem{afk}
A.~Ashtekar, S.~Fairhurst and B.~Krishnan, ``Isolated horizons:
Hamiltonian evolution and the first law,'' Phys.\ Rev.\ D {\bf
62}, 104025 (2000) [arXiv:gr-qc/0005083].

\bibitem{cns} A.~Corichi and D.~Sudarsky,
``Mass of colored black holes,'' Phys.\ Rev.\ D {\bf 61}, 101501
(2000) [arXiv:gr-qc/9912032];
A.~Corichi, U.~Nucamendi and D.~Sudarsky, ``Einstein-Yang-Mills
isolated horizons: Phase space, mechanics, hair and conjectures,''
Phys.\ Rev.\ D {\bf 62}, 044046 (2000) [arXiv:gr-qc/0002078];
A.~Corichi, U.~Nucamendi and D.~Sudarsky, ``A mass formula for EYM
solitons,'' Phys.\ Rev.\ D {\bf 64}, 107501 (2001)
[arXiv:gr-qc/0106084].


\bibitem{acs}
A.~Ashtekar, A.~Corichi and D.~Sudarsky, ``Hairy black holes,
horizon mass and solitons,'' Class.\ Quant.\ Grav.\  {\bf 18}, 919
(2001) [arXiv:gr-qc/0011081].

\bibitem{gibbons} G.W. Gibbons and S.W. Hawking, ``Action Integrals
And Partition Functions In Quantum Gravity,'' Phys.\ Rev.\ D {\bf
15}, 2752 (1977).


\bibitem{Btz}
M.~Banados, C.~Teitelboim and J.~Zanelli, ``Black Hole Entropy And
The Dimensional Continuation Of The Gauss-Bonnet Theorem,'' Phys.\
Rev.\ Lett.\  {\bf 72}, 957 (1994)

\bibitem{HT} M.~Henneaux and C.~Teitelboim,
``Asymptotically Anti-De Sitter Spaces,'' Commun.\ Math.\ Phys.\
{\bf 98}, 391 (1985);
J.~D.~Brown and M.~Henneaux, ``Central Charges In The Canonical
Realization Of Asymptotic Symmetries: An Example From
Three-Dimensional Gravity,'' Commun.\ Math.\ Phys.\  {\bf 104},
207 (1986).

\bibitem{deser}
S.~Deser and R.~Jackiw, ``Three-Dimensional Cosmological Gravity:
Dynamics Of Constant Curvature,'' Annals Phys.\  {\bf 153}, 405
(1984).

\bibitem{ih2} A. Ashtekar, O. Dreyer and J.~Wisniewsk, ``Isolated
horizons in $2+1$ gravity", Adv. Theor. Math. Phys. {\bf 6}, 507
2003. arxiv:gr-qc/0206024.

\bibitem{myung} Y.S. Myung, ``Entropy of the three-dimensional
Schwarzschild-de Sitter black hole", Mod. Phys. Lett. {\bf A16},
2353 (2001).

\bibitem{strom}
M.~Spradlin, A.~Strominger and A.~Volovich, ``Les Houches lectures
on de Sitter space,'' in Les Houches 2001, Gravity, gauge theories
and strings* 423-453 [arXiv:hep-th/0110007].

\bibitem{cg1} A.~Corichi and A.~Gomberoff,
``On a spacetime duality in 2+1 gravity,'' Class.\ Quant.\ Grav.\
{\bf 16}, 3579 (1999) [arXiv:gr-qc/9906078].


\bibitem{nairai} H. Nariai, ``On a New Cosmological Solution to Einstein´s
Field Equations of Gravity". The Science Reports of the Tohoku
University, Series I, {\bf No.1}, (1951). (reprinted in Gen. Rel.
Grav. {\bf 31}, 963 (1999)).

\bibitem{Podolsky:1999ts}
J.~Podolsky, ``The structure of the extreme Schwarzschild-de
Sitter space-time,'' Gen.\ Rel.\ Grav.\  {\bf 31}, 1703 (1999)
[arXiv:gr-qc/9910029].

\bibitem{bousso} R.~Bousso and S.~W.~Hawking,
``Pair creation of black holes during inflation,'' Phys.\ Rev.\ D
{\bf 54}, 6312 (1996) [arXiv:gr-qc/9606052];
R.~Bousso and S.~W.~Hawking, ``(Anti-)evaporation of
Schwarzschild-de Sitter black holes,'' Phys.\ Rev.\ D {\bf 57},
2436 (1998) [arXiv:hep-th/9709224].

\bibitem{asht-das} A.~Ashtekar and S.~Das,
``Asymptotically anti-de Sitter space-times: Conserved
quantities,'' Class.\ Quant.\ Grav.\  {\bf 17}, L17 (2000)
[arXiv:hep-th/9911230].


\bibitem{penrose} R. Penrose, ``Naked Singularities," N. Y. Acad.
Sci. {\bf 224}, 125-134 (1973).


\bibitem{vafa} A.\ Strominger and C.\ Vafa, Microscopic origin
of the Bekenstein-Hawking entropy, {\it Phys.\ Lett.} {\bf B379},
99-104 (1996); J.\ Maldacena and A.\ Strominger, Statistical
entropy of four-dimensional extremal black holes, {\it Phys.\
Rev.\ Lett.} {\bf 77} 428-429 (1996); G. Horowitz, Quantum States
of Black Holes, in {\it Black Holes and Relativistic Stars}, R.
Wald Ed., Chicago University Press, 1998.

\bibitem{boussorev} R.~Bousso,
``Adventures in de Sitter space,'' arXiv:hep-th/0205177.

\bibitem{boussoD} R.~Bousso,
``Bekenstein bounds in de Sitter and flat space,'' JHEP {\bf
0104}, 035 (2001) [arXiv:hep-th/0012052].

\bibitem{beken} J.~D.~Bekenstein,
``A Universal Upper Bound On The Entropy To Energy Ratio For
Bounded Systems,'' Phys.\ Rev.\ D {\bf 23}, 287 (1981).

\bibitem{holog} R.~Bousso,
``The holographic principle,'' Rev.\ Mod.\ Phys.\  {\bf 74}, 825
(2002) [arXiv:hep-th/0203101].

\bibitem{marolf} R.~Bousso, E.~E.~Flanagan and D.~Marolf,
``Simple sufficient conditions for the generalized covariant
entropy bound,'' Phys.\ Rev.\ D {\bf 68}, 064001 (2003)
[arXiv:hep-th/0305149];
R.~Bousso, ``Light-sheets and Bekenstein's bound,'' Phys.\ Rev.\
Lett.\  {\bf 90}, 121302 (2003) [arXiv:hep-th/0210295].

\bibitem{Clarkson} R.~Clarkson, A.~M.~Ghezelbash and R.~B.~Mann,
``Mass, Action And Entropy Of Taub-Bolt-Ds Spacetimes,'' Phys.\
Rev.\ Lett.\  {\bf 91}, 061301 (2003) [arXiv:hep-th/0304097];
R.~Clarkson, A.~M.~Ghezelbash and R.~B.~Mann, ``Entropic N-bound
and maximal mass conjectures violation in four dimensional
Taub-Bolt(NUT)-dS spacetimes,'' Nucl.\ Phys.\ B {\bf 674}, 329
(2003) [arXiv:hep-th/0307059].


\bibitem{kunz} B.~Kleihaus, J.~Kunz, A.~Sood and M.~Wirschins,
``Horizon properties of Einstein-Yang-Mills black hole,'' Phys.\
Rev.\ D {\bf 65}, 061502 (2002) [arXiv:gr-qc/0110084].


\bibitem{higgs} B.~Hartmann, B.~Kleihaus and J.~Kunz,
``Axially symmetric monopoles and black holes in
Einstein-Yang-Mills-Higgs theory,'' Phys.\ Rev.\ D {\bf 65},
024027 (2002) [arXiv:hep-th/0108129].

\bibitem{nora} N.~Breton,
``Born-Infeld black hole in the isolated horizon framework,''
Phys. Rev. {\bf D67}, 124004, (2003). arXiv:hep-th/0301254.

\bibitem{ac:nonmin} A. Ashtekar, A. Corichi and D. Sudarsky,
``Non-minimally coupled scalar fields and isolated horizons,"
Class. Quantum Grav. {\bf 20}, 3413-3425 (2003)
[arXiv:gr-qc/0305044];
A. Ashtekar and A. Corichi, ``Non-minimal couplings, quantum
geometry and black hole entropy", Class. Quantum. Grav. {\bf 20},
 4473-4484 (2003) [arXiv:gr-qc/0305082].

\bibitem{dynamical}
A.~Ashtekar and B.~Krishnan, ``Dynamical horizons: Energy, angular
momentum, fluxes and balance laws,'' Phys.\ Rev.\ Lett.\  {\bf
89}, 261101 (2002) [arXiv:gr-qc/0207080];
A.~Ashtekar and B.~Krishnan, ``Dynamical horizons and their
properties,'' arXiv:gr-qc/0308033;
A.~Ashtekar, ``How black holes grow,'' arXiv:gr-qc/0306115.

\bibitem{ds/cft} A.~Strominger,
``The dS/CFT correspondence,'' JHEP {\bf 0110}, 034 (2001)
[arXiv:hep-th/0106113].

\bibitem{verlinde}
R.~G.~Cai, ``Cardy-Verlinde formula and asymptotically de Sitter
spaces,'' Phys.\ Lett.\ B {\bf 525}, 331 (2002)
[arXiv:hep-th/0111093].
R.~G.~Cai, ``Cardy-Verlinde formula and thermodynamics of black
holes in de Sitter  spaces,'' Nucl.\ Phys.\ B {\bf 628}, 375
(2002) [arXiv:hep-th/0112253].



\end{thebibliography}
\end{document}